\documentclass[preprint,showkeys,amsmath,amssymb,superscriptaddress,nofootinbib]{revtex4}

\usepackage{bm}

\usepackage[pdftex]{graphicx,color}
\usepackage{epstopdf}
\usepackage[bookmarks=true,bookmarksnumbered=true,bookmarkstype=toc]{hyperref} 
\usepackage{units}

\newcommand{\lsim}{\raise0.3ex\hbox{$\;<$\kern-0.75em\raise-1.1ex\hbox{$\sim\;$}}}
\newcommand{\gsim}{\raise0.3ex\hbox{$\;>$\kern-0.75em\raise-1.1ex\hbox{$\sim\;$}}}
\usepackage{mathtools}

\usepackage{color}

\definecolor{green}{cmyk}{1,0,1,0}

\definecolor{pink}{cmyk}{0,0.5,0,0}

\definecolor{pastelpink}{cmyk}{0,0.25,0,0}

\definecolor{softpink}{cmyk}{0,0.125,0,0}

\definecolor{purple}{cmyk}{0.5,1.0,0.1,0}

\definecolor{violet}{cmyk}{0.75,1,0.25,0}

\usepackage{ulem}  

\newcommand{\Begr}{${^8}$Be }                

\usepackage[utf8]{inputenc}

\preprint{UME-PP-005}
\preprint{EPHOU-16-019}

\begin{document}

\title{Atomki anomaly and dark matter in a radiative seesaw model with gauged $B-L$ symmetry}

\author{Osamu Seto}
\email{seto@particle.sci.hokudai.ac.jp}
\affiliation{Institute for International Collaboration, Hokkaido University, Sapporo 060-0815, Japan}
\affiliation{Department of Physics, Hokkaido University, Sapporo 060-0810, Japan}

\author{Takashi Shimomura}
\email{shimomura@cc.miyazaki-u.ac.jp}
\affiliation{Faculty of Education, Miyazaki University, Miyazaki, 889-2192, Japan}

\begin{abstract}
Motivated by recently reported anomalies in a decay of an excited state of beryllium by the Atomki Collaboration, 
we study a radiative 
seesaw model with gauged $B-L$ symmetry and a $Z_2$ parity. 
Assuming that the anomalies originate from the decay of the $B-L$ gauge boson followed by the nuclear decay, 
the mass of the lightest right-handed neutrino or the dark matter candidate can be determined below $10$ GeV. 
We show that for this mass range,  the model can explain the anomalies in the beryllium decay and the relic dark 
matter abundance consistent with neutrino masses. We also predict its
spin-independent cross section in direct detection experiments for this mass range.
\end{abstract}

\date{\today}

\maketitle

\section{Introduction}
The Standard Model (SM) of particle physics has been explaining almost all of experimental results including 
recent LHC data. Despite its enormous success, some phenomena are left unexplained in the SM. 
One of such phenomena is neutrino oscillations, which result in nonzero and tiny neutrino masses as well as flavor 
mixing. Another one is the existence of dark matter (DM). Since neutrinos are massless and no candidates exist in 
the SM, these phenomena are clear evidences of new physics beyond the SM. 

Several mechanisms have been proposed to explain the tininess of neutrino mass. The most popular mechanism is 
the so-called type I seesaw mechanism~\cite{Minkowski:1977sc, Yanagida:1979as, GellMann:1980vs, Mohapatra:1979ia}. 
In the mechanism, right-handed (RH) neutrinos with heavy Majorana masses are 
introduced to the SM, and the tiny neutrino masses can be explained by a suppression of their heavy mass. 
There are other types of seesaw mechanisms, type II~\cite{Konetschny:1977bn, Cheng:1980qt, Schechter:1980gr}, 
type III \cite{Foot:1988aq} and radiative models~\cite{Zee:1980ai, Zee:1985id, Babu:1988ki}.  
In radiative seesaw mechanisms, a discrete parity is generally imposed to the SM so that neutrinos 
can not have tree-level masses or Yukawa interactions (For pioneering works, see e.g. Refs~\cite{Krauss:2002px, Ma:2006km}). 
Then, neutrino masses are generated at loop-level in which new scalars and/or fermions propagate. 
The masses generated are suppressed by the masses of the new particles 
in the loop and a loop factor. Tininess of neutrino masses is explained in this sense. 
In addition to the generation of neutrino mass, the radiative seesaw mechanisms have another virtue. 
The lightest particle with odd parity becomes stable due to the discrete parity. 
Such a stable particle can be a good candidate for the DM. In fact, many radiative seesaw models 
can predict the correct DM abundance. 
Thus, the two phenomena mentioned above can be explained simultaneously. 

Recently, the Atomki Collaboration has reported anomalies in the distributions of the invariant mass and
 the opening angle of an emitted electron-positron pair from the decay of an excited state of
 beryllium (\Begr) into its ground state~\cite{Krasznahorkay:2015iga}. 
They claimed that the deviation from a standard nuclear physics interpretation reaches to $6.8\sigma$, and hence,  
the deviation is probably not a nuclear physics origin. We refer these anomalies to the Atomki anomalies. 
One of the possibilities to explain the anomalies is the subsequent decay of an unknown particle produced in the \Begr decay. 
The Atomki Collaboration assumed a new boson particle with spin-parity $J^\pi = 1^+$ and determined its 
mass as $m = 16.70 \pm 0.35$(stat)$\pm 0.5$(syst) MeV from their data. It is natural to consider that the boson acquires 
such a light mass from a spontaneous breakdown of a symmetry. Then, a fundamental scale of the nature can be determined. 
In \cite{Feng:2016jff, Feng:2016ysn},  the authors showed the Atomki anomalies can be explained by a gauge boson 
in classes of models with gauged {\it baryon} ($B$) and {\it baryon minus lepton} ($B-L$) symmetry.
These gauge symmetries are one of the minimal extensions of the SM, and have been extensively studied in 
terms of various motivations. There are also studies to explain the Atomki anomalies in  
other gauge symmetries~\cite{Gu:2016ege, Neves:2016nek}, with an axial vector~\cite{Kahn:2016vjr} and 
pseudoscalars~\cite{Ellwanger:2016wfe}. Implications on the DM have also been 
studied~\cite{Jia:2016uxs, Kitahara:2016zyb, Chen:2016tdz}. 

In this paper, we study the implications of the Atomki anomalies in a radiative seesaw model with gauged $B-L$ 
symmetry proposed by us~\cite{Kanemura:2011vm}. We find parameter values consistent with experiments 
by taking into account the neutrino mass, the Higgs mass as well as the new boson mass. 
Then, we predict the spin-independent cross section consistent with the dark matter abundance.
Various radiative seesaw mechanism with $Z_2$ parity and gauged $U(1)_{B-L}$ symmetries have been 
proposed~\cite{Li:2010rb, Kanemura:2011vm, Khalil:2011tb, Lindner:2011it, Kanemura:2011mw, Okada:2012np, 
Kanemura:2012zh, Kajiyama:2013zla, Kajiyama:2013rla, Basak:2013cga, Kanemura:2014rpa, Okada:2014nea, 
Wang:2015saa}. 
Our following discussion and results would be applicable
once one tries to address the Atomki anomalies in such a model,
because the required cross section and the mass determine the scale of the $U(1)_{B-L}$ symmetry. 

The rest of this paper is organized as follows. In section~\ref{model}, we explain our model including brief review 
of neutrino masses. We show the interaction Lagrangian of the gauge boson with SM fermions as well as 
constraints for the Atomki anomalies to be explained. Then, parameters and masses consistent with experimental constraints are derived in section~\ref{parameters}. In section~\ref{dark-matter}, the spin-independent cross 
section is predicted for the parameter values derived in section~\ref{parameters}. 
We summarize our study in section~\ref{summary}.

\section{Model} \label{model}
We explain our model proposed in Ref.~\cite{Kanemura:2011vm}. 
The SM is extended by imposing the gauged $U(1)_{B-L}$ symmetry and a $Z_2$ parity, and also 
introducing two scalar particles and
 three right-handed neutrinos, $N_R$. One of the scalar particles, $S$, is a SM singlet and 
responsible for $B-L$ symmetry breaking. The other one, $\eta$, has the same quantum charge as the 
SM Higgs $\Phi$ and is related 
with the generation of neutrino masses. The scalar $\eta$ and the RH neutrinos $N_R$ are $Z_2$ odd while other 
particles are $Z_2$ even. The charge assignment of the particles is summarized in Table~\ref{particle}. Here, $Q^i,~d_R^i,~u_R^i$ and 
$L^i,~e_R^i$ are the left-handed (LH) and the right-handed quarks and leptons, respectively. Latin and Greek indices 
denote generation and flavor of fermions.
\begin{table}
\begin{center}
  \begin{tabular}{|c|cccccc|ccc|}
   \hline
   & $Q^i$ & $d_R^{i}$ & $u_R^{i}$ & $L^i$ & $e_R^i$ & 
   $\Phi$ & $\eta$ & $S$ & $N_{R}^{\alpha}$  \\\hline 
$SU(3)_C$ & $3$ & $3$ & $3$ & $1$ & $1$ & $1$ & $1$ & $1$ & $1$ \\\hline
$SU(2)_W$ & $2$ & $1$ & $1$ & $2$ & $1$ & $2$ & $2$ & $1$ & $1$ \\\hline
$U(1)_{Y}$ & $1/6$ & $-1/3$ & $+2/3$ & $-1/2$ & $-1$ & $1/2$ & $1/2$ & $0$ & $0$ \\\hline
$U(1)_{B-L}$ & $1/3$ & $1/3$ & $1/3$ & $-1$ & $-1$ & $0$ & $0$ & $+2$ & $-1$ \\\hline
$Z_2$   & $+$ & $+$ & $+$ & $+$ & $+$ & $+$ & $-$ & $+$ & $-$ \\ \hline  
   \end{tabular}
\end{center}
  \caption{The charge assignment of fields. }
  \label{particle}
 \end{table}

First, we briefly review the neutrino mass in our model. The interaction Lagrangian for the generation of 
neutrino mass is given by 
\begin{align}
\mathcal{L}_{\mathrm{int}} = \mathcal{L}_N - V(\Phi, \eta, S),
\end{align}
where the Yukawa interactions are given by
\begin{align}
\mathcal{L}_N = g_{i \alpha} \overline{L}^i \tilde{\eta} N^\alpha_R 
- \frac{Y_R^\alpha}{2} S  \overline{(N^\alpha_R)^c} N^\alpha_R + h.c., 
\end{align} 
and the scalar potential is given by
\begin{align}
V(\Phi, \eta, S) &= \mu_1^2 |\Phi|^2 + \mu_2^2 |\eta|^2 + \mu_S^2 |S|^2  +\lambda_1 |\Phi|^4  
   + \lambda_2|\eta|^4 + \lambda_3 |\Phi|^2|\eta|^2 + \lambda_4|\Phi^{\dagger} \eta|^2 \nonumber \\
 & \quad + \frac{\lambda_5}{2} \left[ (\Phi^{\dagger} \eta)^2 + {\rm h.c.} \right] +\lambda_S |S|^4 
  + \tilde{\lambda} |\Phi|^2 |S|^2 + \lambda |\eta|^2|S|^2,
\label{eq:scalar-pot}
\end{align} 
with $\tilde{\eta} = i \sigma_2 \eta^\ast$. 
The summation over repeated indices should be understood.
Because of the $Z_2$ parity, neutrinos can not have Yukawa couplings with the Higgs field $\Phi$; 
instead, they can have those with $\eta$. 
After the Higgs and the scalar $S$ develop vacuum expectation values (VEVs),
\begin{align}
\langle \Phi \rangle = \frac{1}{\sqrt{2}}
\begin{pmatrix}
0 \\
v
\end{pmatrix},
~~
\langle S \rangle = \frac{v_S}{\sqrt{2}},
\end{align}
where $v = 246$ GeV, the masses of neutrinos are generated via a one-loop diagram in which $N_R$ and 
$\eta$ propagate, and expressed as 
\begin{align}
 m_{\nu_L}^{ij} \simeq  
  \frac{\lambda_5}{8 \pi^2} g_{i \alpha} Y_R^\alpha g_{\alpha j}^T  \left(\frac{v}{m_{\eta}}\right)^2  v_S. 
  \label{eq:neutrino-mass}
\end{align}
More details can be found in Ref.~\cite{Kanemura:2011vm}.

Next, we consider the gauge sector. The relevant Lagrangian to explain the Atomki anomalies is 
given by
\begin{align}
\mathcal{L} = \mathcal{L}_{\mathrm{gauge, int}} + \mathcal{L}_{\mathrm{gauge, kin}}, 
\end{align}
where
\begin{subequations}
\begin{align}
\mathcal{L}_{\mathrm{gauge, int}} &= g_1 \hat{B}_\mu J^\mu_1 + g_2 \hat{W}^a_\mu J^{a \mu}_2 
 +\epsilon_X e \hat{X}_\mu J_{X}^\mu, \\
\mathcal{L}_{\mathrm{gauge, kin}} &= -\frac{1}{4} \hat{B}_{\mu \nu} \hat{B}^{\mu \nu} 
               -\frac{1}{4} {\hat{W}}^{a}_{\mu \nu} \hat{W}^{a \mu \nu} 
               -\frac{1}{4} \hat{X}_{\mu \nu} \hat{X}^{\mu \nu}                        
               +\frac{\epsilon}{2} \hat{B}_{\mu \nu} \hat{X}^{\mu \nu}.
\end{align}
\end{subequations}
Here, $\hat{B},~\hat{W}^a, ~\hat{X}$ represent $U(1)_Y$, $SU(2)_W$ and $U(1)_{B-L}$ gauge bosons in the interaction basis,  
in which $a$ being $SU(2)$ index, while $\hat{B}^{\mu \nu},~\hat{W}^{a \mu \nu}, ~\hat{X}^{\mu \nu}$ are their field strengths, respectively. 
The coupling constants and the currents of $U(1)_Y$, $SU(2)_W$ and $U(1)_{B-L}$ are denoted as $g_1,~g_2$ and $\epsilon_X e$, 
and $J^\mu_1,~J^{a \mu}_2$ and 
\begin{align}
J^\mu_X = \frac{1}{3} \overline{u^i} \gamma^\mu u^i  + \frac{1}{3} \overline{d^i} \gamma^\mu d^i 
   - \overline{e^i} \gamma^\mu e^i - \overline{\nu^i} \gamma^\mu \nu^i  - \overline{N_R^i} \gamma^\mu N_R^i .
\end{align}
Note that the gauge coupling constant of $U(1)_{B-L}$ is normalized by the electric charge of 
electron for convenience. The kinetic mixing parameter is denoted as $\epsilon$.

After the electroweak and the $B-L$ symmetries are broken,
 the interaction Lagrangian of the neutral gauge bosons in the mass basis is given as
\begin{subequations}
\begin{align}
\mathcal{L}_{\mathrm{gauge, int}} &= e A_\mu J^\mu_{em} 
         + Z_\mu \bigg[ g_2 (c_\chi - \varepsilon s_W s_\chi) J^{\mu}_{NC}  + \varepsilon c_W s_\chi J^\mu_{em} 
          + \varepsilon_X e s_\chi J^\mu_X \bigg] \nonumber \\
         & \quad + X_\mu \bigg[ \varepsilon_X e  c_\chi J_{X}^\mu  + \varepsilon e c_W c_\chi J^\mu_{em} 
         	-g_2 (s_\chi + \varepsilon s_W c_\chi) J^\mu_{NC} \bigg], \\
\mathcal{L}_{\mathrm{gauge, kin}} &= -\frac{1}{4} F_{\mu \nu} F^{\mu \nu} 
               -\frac{1}{4} Z_{\mu \nu} Z^{\mu \nu} + \frac{1}{2} m_Z^2 Z_\mu Z^\mu 
               -\frac{1}{4} X_{\mu \nu} X^{\mu \nu} + \frac{1}{2} m_X^2 X_\mu X^\mu, 
\end{align}
\end{subequations}
where $A_\mu (F_{\mu\nu})$ and $Z_\mu (Z_{\mu \nu})$ represent the SM photon and the $Z$ boson (and their field strengths), 
respectively. The currents $J_{em}^\mu$ and $J_{NC}^\mu$ are the same as those in the SM. 
The dimensionless parameters $\varepsilon$ and $\varepsilon_X$ are defined as
\begin{align}
\varepsilon = \epsilon r,~~~~\varepsilon_X = \epsilon_X r,
\end{align}
with $r = (1-\epsilon^2)^{-1/2}$.
The weak mixing angle is denoted as $s_W = \sin\theta_W~(c_W=\cos\theta_W)$, and the mixing angle of 
the gauge bosons due to the kinetic mixing, 
$s_\chi = \sin\chi~(c_\chi = \cos\chi)$, is defined by
\begin{align}
\tan 2\chi = \frac{- m_{\hat{Z}}^2 q}{(1- q^2) m_{\hat{Z}}^2  - m_{\hat{X}}^2 r^2  },
\end{align}
with $q= - \varepsilon s_W$, and 
\begin{subequations}
\begin{align}
m_{\hat{Z}} &= \frac{1}{2}\sqrt{g_1^2 + g_2^2} v, \\
m_{\hat{X}} &= 2 \epsilon_X e v_s.
\end{align}
\end{subequations}
The masses of the gauge bosons are given as 
\begin{subequations}
\begin{align}
m_{Z}^2 &= \frac{1}{2} \left[ m_{\hat{Z}}^2 (1+q^2) + m_{\hat{X}}^2 r^2 + \sqrt{D} \right], \\
m_{X}^2 &= \frac{1}{2} \left[ m_{\hat{Z}}^2 (1+q^2) - m_{\hat{X}}^2 r^2 + \sqrt{D} \right], \\
D&= ( m_{\hat{Z}}^2 (1+q^2) + m_{\hat{X}}^2 r^2 )^2 - 4 m_{\hat{Z}}^2 m_{\hat{X}}^2 r^2.
\end{align}
\end{subequations}
For $|\varepsilon|,~|\varepsilon_X| \ll 1$, the mixing angle can be approximated 
\begin{align}
s_\chi \simeq - \epsilon s_W, ~~~~c_\chi \simeq 1.
\end{align}
Then, the interaction Lagrangian and the gauge boson masses are given by
\begin{subequations}
\begin{align}
\mathcal{L}_{\mathrm{gauge, int}} &= e A_\mu J^\mu_{em}  + g_2 Z_\mu  J^{\mu}_{NC}  
         + e X_\mu \bigg[ \varepsilon_X   J_{X}^\mu  + \varepsilon c_W  J^\mu_{em} \bigg]
         + \mathcal{O}(\epsilon^2,\epsilon\epsilon_X), \label{eq:gauge-int} \\
m_{Z}^2 &\simeq  m_{\hat{Z}}^2 + \mathcal{O}(\epsilon^2), \\
m_{X}^2 &\simeq m_{\hat{X}}^2 + \mathcal{O}(\epsilon^2). \label{eq:mass-x-boson}
\end{align}
\end{subequations}
>From Eq.~\eqref{eq:gauge-int}, the resulting coupling constants of the SM fermions to the $X$ boson are 
\begin{subequations}
\begin{align}
\varepsilon_u &= \frac{1}{3} \varepsilon_X + \frac{2}{3} \varepsilon c_W, \\
\varepsilon_d &= \frac{1}{3} \varepsilon_X - \frac{1}{3} \varepsilon c_W, \\
\varepsilon_\nu &= - \varepsilon_X,  \\
\varepsilon_e &= - \varepsilon_X - \varepsilon c_W.
\end{align}
\end{subequations}
The coupling constants of an up and a down quark can be translated into those of a proton and a neutron as 
\begin{align}
\varepsilon_p = 2 \varepsilon_u + \varepsilon_d,~~~\varepsilon_n = \varepsilon_u + 2 \varepsilon_d.
\end{align}
To explain the Atomki anomalies, $|\varepsilon_n|$ and $|\varepsilon_p|$ are required to satisfy~\cite{Feng:2016ysn}
\begin{subequations}
\begin{align}
|\varepsilon_n| &= (2-10) \times 10^{-3}, \\
|\varepsilon_p| &\lesssim 1.2 \times 10^{-3}.
\end{align}
\end{subequations}
On the other hand, these coupling constants are constrained by several experiments, i.e. the dark photon searches in neutral pion decays, beam dump searches, neutrino-electron scatterings. 
The constraints as well as the signal requirements are summarized as~\cite{Feng:2016ysn}
\begin{subequations}
\begin{align}
|\varepsilon_n| & = |\varepsilon_X| =  (2-10) \times 10^{-3},  \label{eq:eps-x-bound} \\
|\varepsilon_p| & = |\varepsilon_X - \varepsilon c_W| \lesssim 1.2 \times 10^{-3} \\
|\varepsilon_e| & = (0.2 - 1.4) \times 10^{-3} \\
\sqrt{|\varepsilon_e  \varepsilon_\nu|} & \lesssim 3 \times 10^{-4}.
\end{align}
\end{subequations}
Note that in our model, $\varepsilon_\nu = - \varepsilon_n$ and hence the above experimental constraints 
are not satisfied. However, it is possible to evade the constraint by further extensions. 
One of such successful extensions is to introduce pairs of vectorlike leptons~\cite{Feng:2016ysn} whose SM 
gauge charges are the same while the $B-L$ charge is opposite. 
Because of the opposite $B-L$ charge, the mixing between the LH neutrinos and the vectorlike neutrinos can 
suppress the lightest neutrino coupling to $X$ so that the constraints can be satisfied. This extension can be 
applied to our model and make the lightest neutrino to be neutralized to the $X$ boson. 
For concreteness, we consider one pair of the vectorlike leptons $L^4_L =(\nu^4_L, e^4_L)^T,~L^4_R=(\nu^4_R, e^4_R)^T$ 
and $E^4_L,~E^4_R$ that are $SU(2)_W$ doublet and singlet, respectively. We assign the even parity of $Z_2$ to these 
leptons. Then, the mass term of the neutrinos after the symmetry breaking is given by
\begin{align}
\mathcal{L}_{\mathrm{mass}} = -\frac{1}{2}m_M \overline{N_R^c} N_R 
  - M_{4L} \overline{{\nu^4_R}} \ \nu_L - M_L \overline{\nu^4_L} \ \nu^4_R  + h.c., \label{eq:new-mass-lag}
\end{align}
where $m_M$ and $M_{4L}$ are proportional to $v_S$, and $M_L$ is a Dirac mass. 
It is important to note that the RH neutrinos can not mix with the other neutrinos due to the $Z_2$ parity and 
hence, be taken as mass eigenstates.
The second and the third terms of Eq.\eqref{eq:new-mass-lag} can be casted 
as $\overline{\psi^\nu_L} M_\nu \psi^\nu_R + h.c.$ where
\begin{subequations}
\begin{align}
M_\nu &= 
\begin{pmatrix}
0 & M_{4L} \\
0 & M_L
\end{pmatrix}, \label{eq:mass-new-vector} \\
\psi^\nu_{L} &= 
\begin{pmatrix}
\nu_L \\
\nu^4_L
\end{pmatrix},\\
\psi^\nu_{R} &= 
\begin{pmatrix}
N_R \\
\nu^4_R
\end{pmatrix}.
\end{align}
\end{subequations}
Diagonalizing Eq.\eqref{eq:mass-new-vector}, we obtain the one Dirac state 
with mass $\sqrt{M_L^2 + M_{4L}^2}$ and one massless state. The latter state 
can be express as
\begin{align}
\frac{1}{\sqrt{M_L^2 + M_{4L}^2}} (-M_L \nu + M_{4L} \nu^4_L),
\end{align}
and its coupling to the $X$ boson is given by
\begin{align}
\epsilon_\nu = -\epsilon_X \cos 2\theta_{\nu},
\end{align}
where $\tan\theta_{\nu} = M_{4L}/M_L$. Thus the lightest neutrino can be neutralized by 
choosing $M_{4L} \simeq M_L$. Generalization of $N$ pairs of the vectorlike leptons is straightforward and 
that allows new vector lepton to be heavier while the lightest neutrino is kept neutralized.

\section{Parameter Values} \label{parameters}
In this section, we show parameter values in the model taking into account the $X$ gauge boson mass and 
couplings. 

Firstly, the VEV of $S$ can be determined since the mass of the gauge boson \eqref{eq:mass-x-boson} should be 
$m_X = 16.70 \pm 0.35$(stat)$\pm 0.5$(syst) MeV~\cite{Krasznahorkay:2015iga},
\begin{align}
v_s = 13.78 \left( \frac{2 \times 10^{-3}}{|\varepsilon_X|} \right)~\mathrm{GeV}, \label{eq:vev-s}
\end{align}
for the central value of $m_X$. We normalize $|\varepsilon_X|$ by $2 \times 10^{-3}$ as a reference value, 
which corresponds to the lower bound in Eq.~\eqref{eq:eps-x-bound}. 
On the other hand, the Higgs mass is given roughly by $\sqrt{2 \lambda_1} v$
 where $\lambda_1$ is the quartic coupling in Eq.~\eqref{eq:scalar-pot}. 
Therefore, the quartic coupling $\lambda_1$ must be $0.130$ to 
reproduce the Higgs mass $125$ GeV. Using Eq.~\eqref{eq:vev-s}, the ratio of the VEVs, $\tan\beta$, and 
the mixing angle between $Z_2$ even scalars, $\alpha$, defined in Ref.~\cite{Kanemura:2011vm} are expressed as
\begin{subequations}
\begin{align}
\tan\beta &=  5.60 \times 10^{-2} \left( \frac{2 \times 10^{-3}}{|\varepsilon_X|} \right),\\
\alpha &\simeq 10^{-4} \left( \frac{\tilde{\lambda}}{4.65 \times 10^{-4}} \right) 
 \left( \frac{0.130}{\lambda_1} \right)  \left( \frac{2 \times 10^{-3}}{|\varepsilon_X|} \right),
\end{align}
\end{subequations}
where $\tilde{\lambda}$ and $\lambda_1$ are normalized by reference values, respectively. 

Then, the masses of the $Z_2$ even lighter scalar $H$ and $N_R$ are 
parametrized by
\begin{subequations}
\begin{align}
m_H &= 19.5 \left( \frac{ \lambda_s }{ 1 } \right)^{1/2} \left( \frac{2 \times 10^{-3}}{|\varepsilon_X|} \right)~\mathrm{GeV},
\label{mass:H} \\
m_{N_R} &= 9.75 \left( \frac{ Y_R }{1} \right) \left( \frac{2 \times 10^{-3}}{|\varepsilon_X|} \right)~\mathrm{GeV},
\label{mass:NR} 
\end{align}
\label{eq:masses}
\end{subequations}
where $\lambda_s$ and $Y_R$ is a quartic coupling in the scalar potential and the couplings of RH neutrinos to $S$, 
respectively. One can see from Eqs.~\eqref{eq:masses} that the masses of $H$ and $N_R$ are less 
than $20$ and $10$ GeV, respectively, when we require $\lambda_s,Y_R \leq 1$. 
These mass ranges are a direct consequence of the light gauge boson because the masses are proportional to 
$v_S$.
Since the other $Z_2$ odd particle, $\eta$, should be heavier than TeV to give tiny neutrino masses, the 
lightest RH neutrino is the DM candidate in our model. 

In the end, the LH neutrino masses given in Eq.~\eqref{eq:neutrino-mass} are parametrized as
\begin{align}
m_{\nu_L} \sim 0.10 \left( \frac{\lambda_5 g_{i \alpha}^2}{5.7 \times 10^{-9} } \right)
     \left( \frac{Y_R}{1} \right) \left( \frac{v/m_{\eta}}{0.1} \right) 
     \left( \frac{2 \times 10^{-3}}{|\varepsilon_X|} \right)~\mathrm{eV}.
\end{align}
Adjusting $\lambda_5$ and $g_{i \alpha}$,  the LH neutrino masses can be taken to the correct 
order, $0.1$ eV, without conflicting other constraints. 
As shown in the next section, direct detection searches of the DM constrains $\alpha$ to be as small as $\mathcal{O}(10^{-4})$, 
and hence $\tilde{\lambda}$ should be smaller than $\mathcal{O}(10^{-4})$.
This fact implies that $H$ and $N_R$ almost decouple from the SM sector because $H$ consists mainly of $S$. 
The SM Higgs boson $h$ can decay into a pair of three light particles. 
However, the decay widths of $h$ into $H H$, $N_R N_R$, and $X X$ pairs are 
\begin{subequations}
\begin{align}
\Gamma(h \rightarrow N_R N_R) &= 6.21 \times 10^{-7}~\mathrm{MeV}, \\
\Gamma(h \rightarrow H H) &= 1.03 \times 10^{-3}~\mathrm{MeV}, \\
\Gamma(h \rightarrow X X) &= 1.02 \times 10^{-3}~\mathrm{MeV},
\end{align}
\end{subequations}
for the above reference values. Therefore the contribution to the invisible Higgs decay width is 
negligible. 

\section{Dark Matter} \label{dark-matter}

The thermal relic abundance of the lightest right-handed neutrino $N_R$ dark matter
 is obtained by integrating the Boltzmann equation for its number density $n$,
\begin{equation}
 \frac{d n }{dt}+3\left(\frac{\dot{a}(t)}{a(t)}\right) n =-\langle\sigma v\rangle ( n^2 - n_{\rm EQ}^2),
\end{equation}
 where $a(t)$ is the scale factor of the expanding Universe,
 the dot denotes the derivative with respect to time,
 $\langle\sigma v\rangle$ is the thermal averaged annihilation cross section
 times relative velocity, and $n_{\rm EQ}$ is the dark matter number density in thermal equilibrium, 
 respectively. 
 
\begin{figure}[t]
\begin{center}
\begin{tabular}{cccc}
\includegraphics[width=6.5cm]{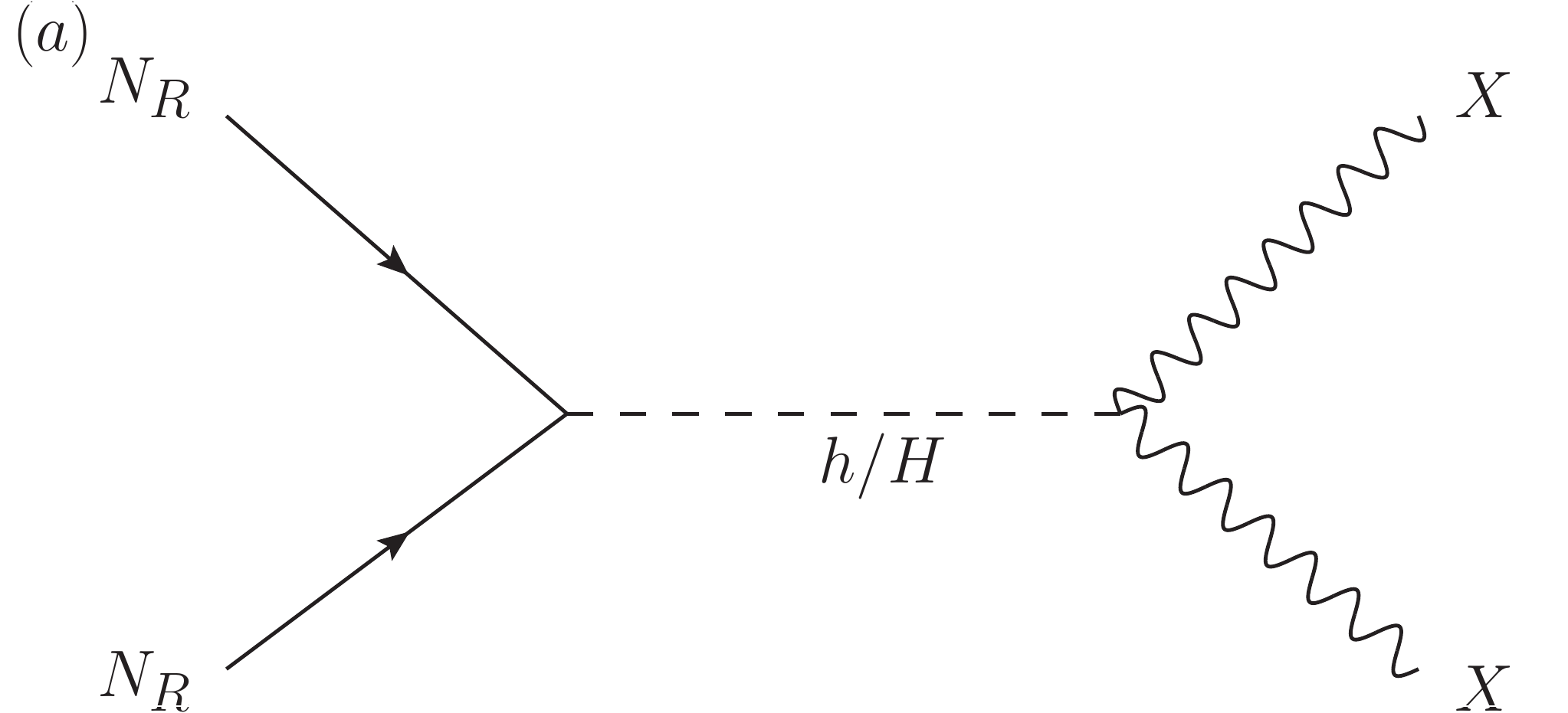}
& &
&
\includegraphics[width=6.5cm]{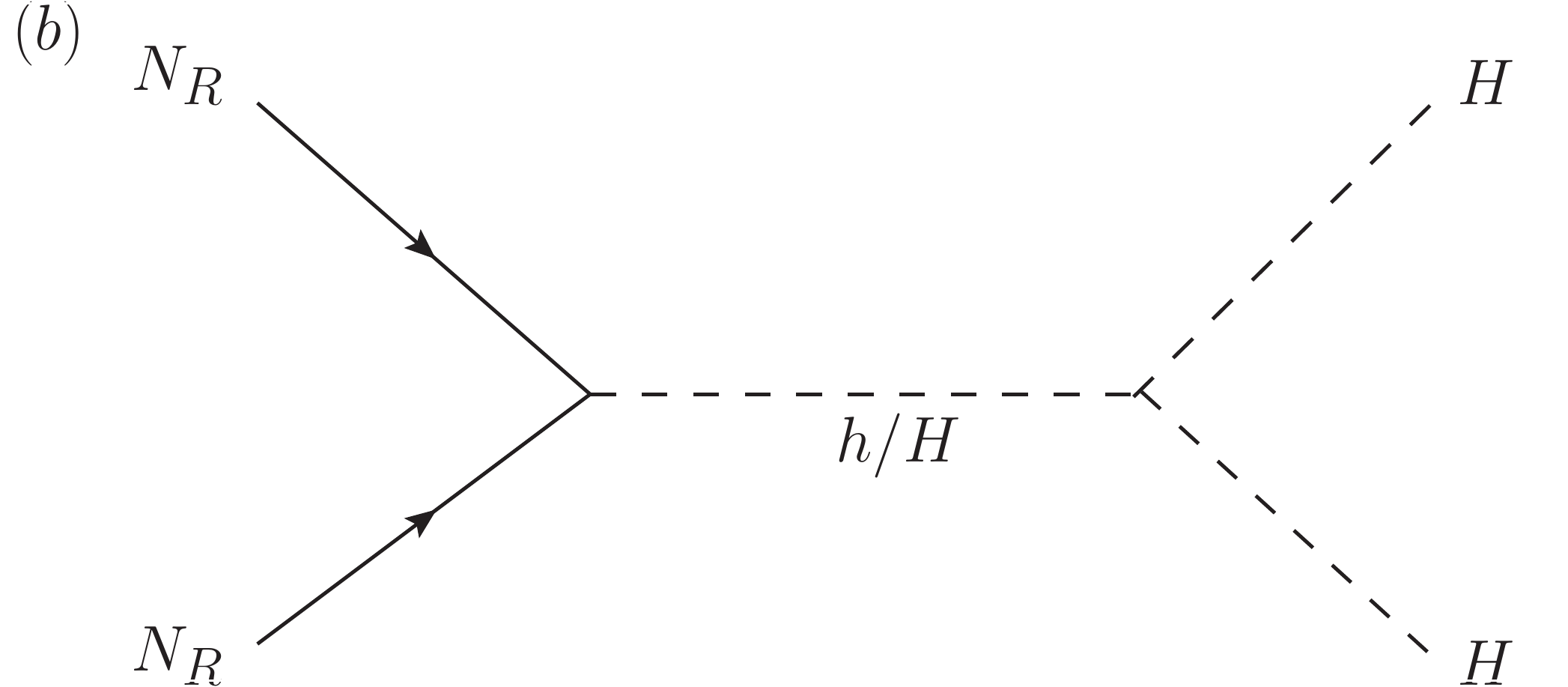} \\
& & &  \\
& & & \\
\includegraphics[width=6.5cm]{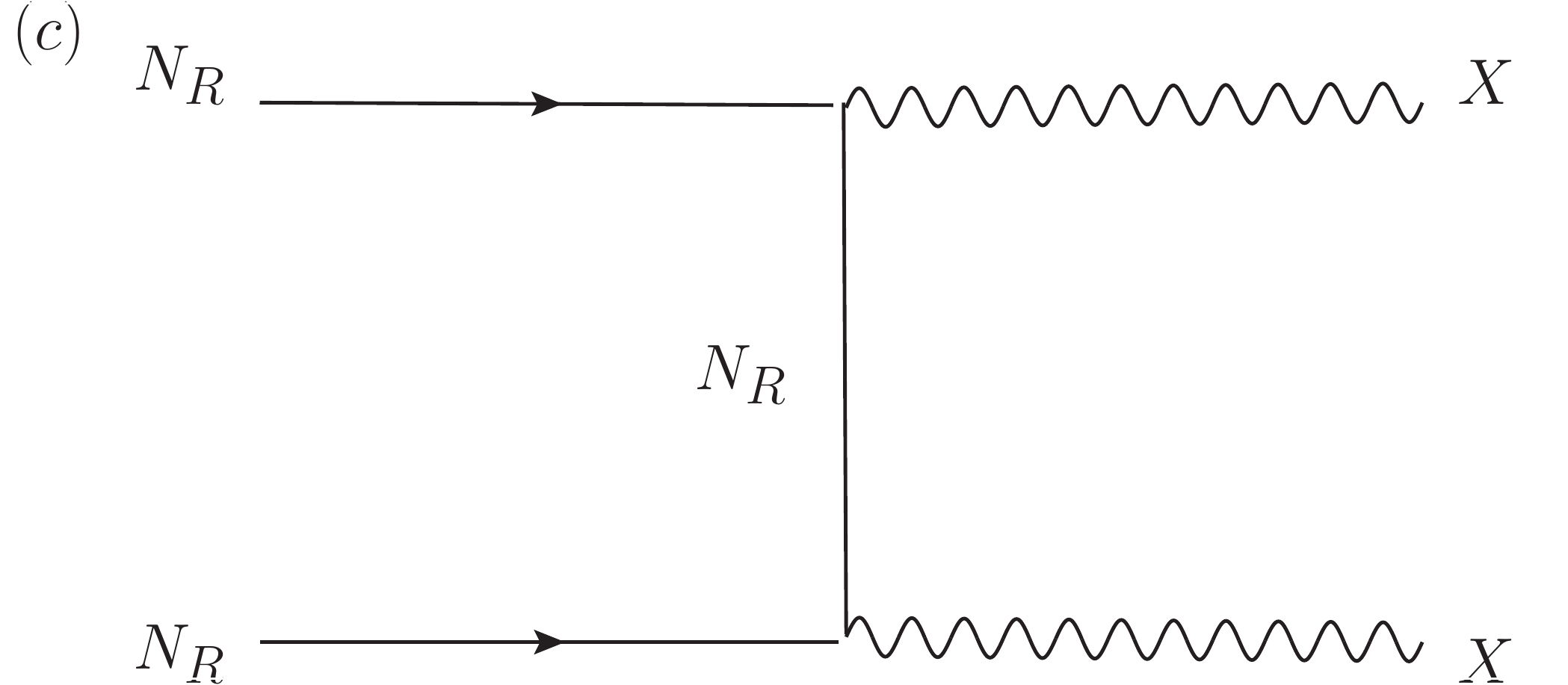}
& &
&
\includegraphics[width=6.5cm]{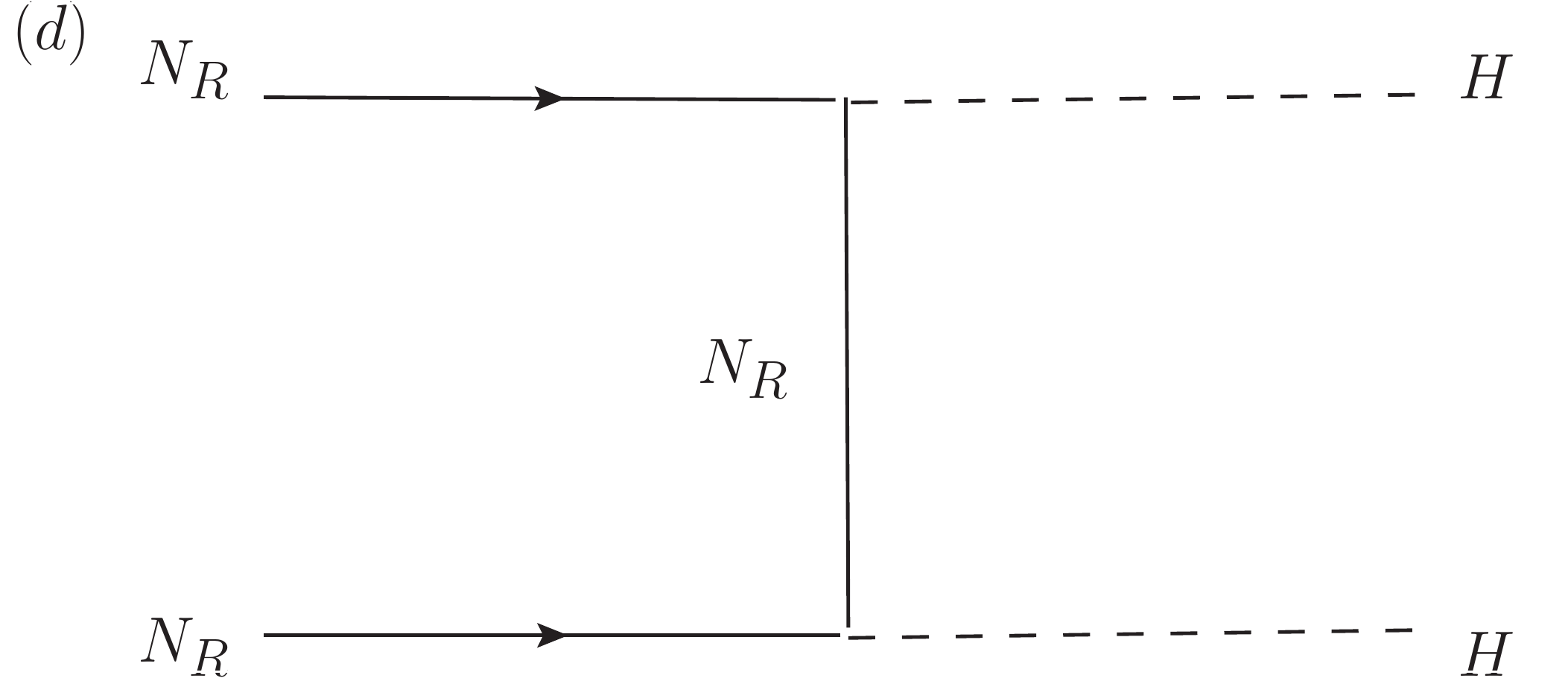} 
\end{tabular}
\end{center}
\caption{ 
The RH neutrino annihilation channels into $XX$ and $HH$.
}
\label{fig:DM-anni-diagrams}
\end{figure}
 
\begin{figure}[t]
\begin{center}
\includegraphics[width=12cm, height=8cm]{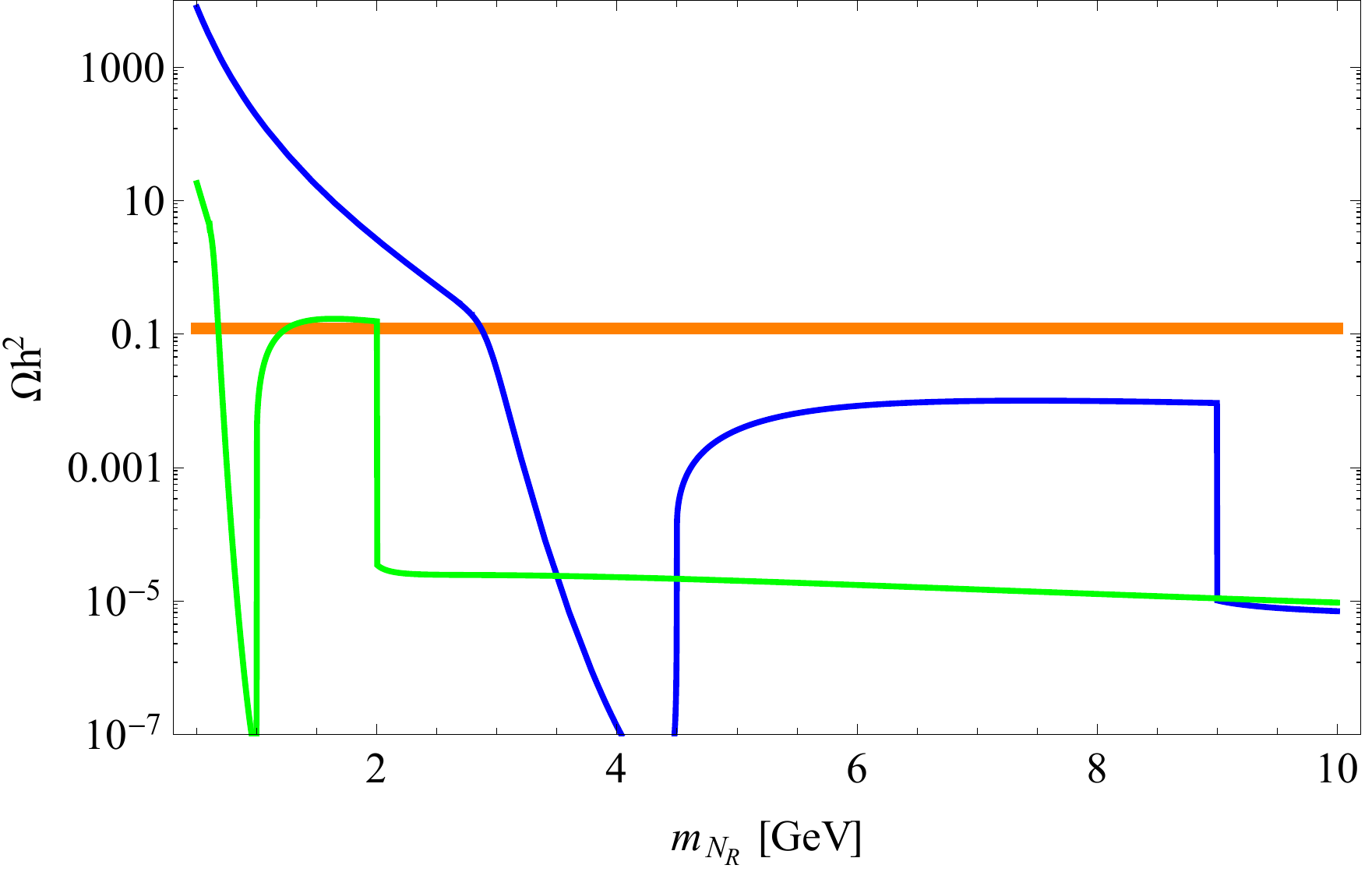}
\end{center}
\caption{ 
 The right-handed neutrino dark matter thermal relic abundance.
 The blue and green lines are
 for $m_H = 9$ GeV and $2$ GeV, respectively.
 The orange line indicates $\Omega_{DM}h^2 \simeq 0.12$ as measured by Planck satellite.
 }
\label{Fig:Omega}
\end{figure}

 The lightest RH neutrino dominantly annihilates into pairs of the $X$ boson and the second Higgs boson $H$ via 
s-channels $H$ exchange shown in Fig.~\ref{fig:DM-anni-diagrams}(a) and 1(b).
Annihilation cross section into $X$ and $H$ bosons pair through t-(u-) channels of the $N_R$ exchange 
in Fig.~\ref{fig:DM-anni-diagrams}(c) and 1(d) 
 is $10^5$ times smaller than that of the s-channel $H$ exchange and hence, negligible.
Other annihilation modes into the SM fermions
 through the s-channel exchange of the Higgs bosons ($h$ and $H$) and the $X$ gauge boson~\cite{Okada:2010wd}
 are less important, we have included those modes in our numerical calculation nevertheless.
We note formula for those annihilation modes in the Appendix for information.
The dominant annihilation mode is  $N_R N_R \rightarrow X X$ for a large parameter region. 
With suggested couplings constant values shown in Eq.~(\ref{eq:eps-x-bound}), 
 we obtain
\begin{equation}
 m_{DM} \simeq 3.4~\textrm{GeV},
\end{equation}
 in a heavy $H$ cases ($m_H \gtrsim 9$ GeV)
 and which is the maximum dark matter mass in our model.
Figure~\ref{Fig:Omega} shows the thermal relic abundance of the lightest right-handed neutrino $\Omega_{N_R} h^2$ 
in terms of $m_{N_R}$.
The orange line indicates $\Omega_{DM}h^2 \simeq 0.12$ as measured by the Planck satellite~\cite{Ade:2015xua}.
The blue and green curves are for $m_H = 9$ GeV and $2$ GeV, respectively, as reference values.
For light $m_H$,
 lighter DM mass regions also become viable due to features depend on $m_H$.
Although typical dark matter abundance is large for $m_{DM} < 3.4$ GeV,
 even in such a region, 
 the relic abundance is significantly reduced by the resonant annihilation for $m_{N_R} = m_H/2$ and
 can meet with its observed value.
This annihilation appears as the deep and narrow gaps in the abundance, which can be seen in the figure. 
The other characteristic appears at the $m_{N_R} = m_H$
 where the annihilation channel into a $HH$ pair is kinematically open, and it dominates the annihilation  
 cross section. The sudden decrease of the relic abundance on the left side is explained by this mode. 

Thus, one can understand from the figure that the dark matter abundance can be reproduced for the two 
cases of $m_{N_R}$:  
(1) at just below the threshold of the $H$ pair annihilation and (2) at both sides of the $H$ resonance or at the off resonance. 
In the former case, the dark matter mass is determined by
\begin{equation}
 m_{DM} \simeq m_H.
\label{thresh}
\end{equation}
In the Fig.~\ref{Fig:Omega}, the dark matter mass is obtained as  
\begin{align}
m_{DM} = 2~\mathrm{GeV},
\end{align}
for $m_H = 2$ GeV. As $m_{N_R}$ as well as $m_H$ are heavy, the annihilation cross section becomes larger and hence 
the dark matter abundance can not be explained for $m_H =9$ GeV.
In the latter case, the dark matter mass is given by 
\begin{equation}
 m_{DM}^\pm= \frac{m_H}{2} \pm \delta m,
\label{reso}
\end{equation}
where the mass difference $\delta m$ is typically from $0.5$ GeV ($m_H = 2$ GeV) to $1$ GeV ($m_H = 9$ GeV).
For $m_H = 9$ GeV, the dark matter mass is determined as 
\begin{equation}
 m_{DM}^- \sim 3.4~\textrm{GeV},
\label{off-reso}
\end{equation}
while for $m_H = 2$ GeV, 
\begin{align}
m_{DM}^\pm = 0.5~\mathrm{or}~1.5~\mathrm{GeV}.
\end{align}
We note that the results are independent from $\sin\alpha$ as its main annihilation mode are.
One may notice that those annihilation processes have a tiny s-wave component and are dominantly p-wave. 

\begin{figure}[ht]
\begin{center}
\includegraphics[width=8cm, height=8cm]{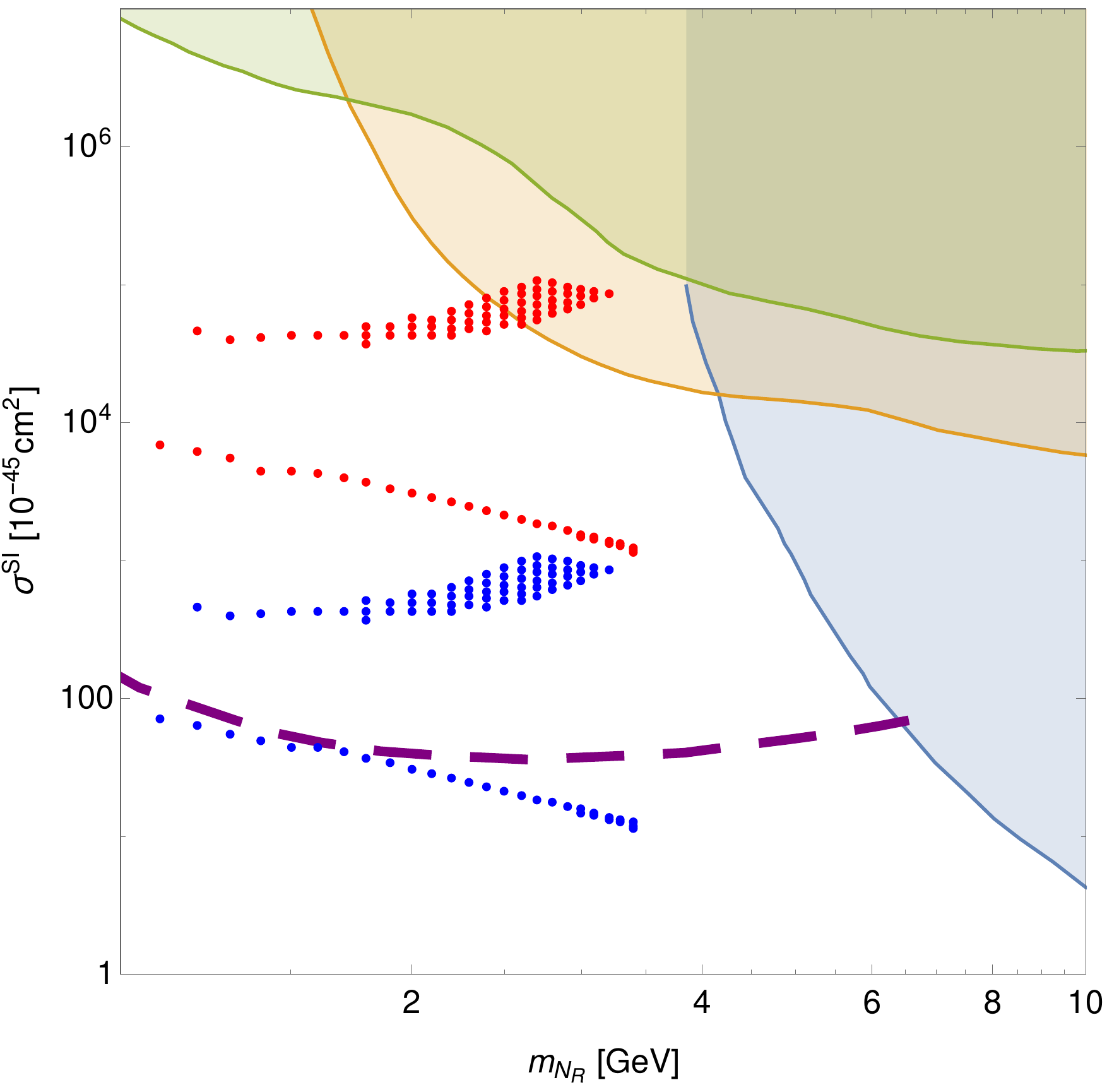}
\end{center}
\caption{ 
 The spin-independent scattering cross section with a proton.
 The red (blue) points are for $\sin\alpha = 1\times10^{-4}$ ($ 1\times10^{-5}$).
 The excluded regions by null results in the CREEST-II, the CDMSlite and the LUX 
 have shading with green, orange and blue, respectively.
 The purple dashed line indicates the expected sensitivity of Ge HV detector in the superCDMS SNOWLAB
 }
\label{Fig:DD}
\end{figure}

This $N_R$ dark matter can be searched
 through the elastic scattering off with a nucleon. 
The spin-independent scattering cross section with a proton through Higgs bosons exchange
 is given by~\cite{Jungman:1995df}
\begin{equation}
\sigma^{\rm SI} = \frac{4}{\pi}
 \left(\frac{m_p m_{N_R}}{m_p + m_{N_R}}\right)^2 f_p^2, 
 \label{sigmaSI}
\end{equation}
 with the proton mass $m_p$ and the effective spin-independent
 coupling between $N_R$ and a proton, $f_p$, which is given as
\begin{equation}
 \frac{f_p}{m_p} = \sum_{q=u,d,s}f_{Tq}^{(p)}\frac{\alpha_q}{m_q} 
  + \frac{2}{27}f_{TG}^{(p)}\sum_{c,b,t}\frac{\alpha_q}{m_q},
\end{equation}
 where $m_q$ is a quark mass, $f_{Tq}^{(p)}$ and $f_{TG}^{(p)}$
 are constants.
The effective vertices between $N_R$ and a quark also
 have been derived in Ref.~\cite{Okada:2010wd} as
\begin{equation}
 \alpha_q = -\frac{m_{N_R} m_q}{v_s v} \sin\alpha \cos\alpha \left(
 \frac{1}{m_h^2 } - \frac{1}{m_H^2} \right) .
 \label{alphaq}
\end{equation}
Figure~\ref{Fig:DD} displays the prediction of $\sigma^{\rm SI}$.
We have searched points satisfying $\Omega_{N_R} h^2 \simeq 0.1$ by 
varying the masses of dark matter $m_{N_R}$ and mediator $m_H$.  
The red (blue) points show the results
 for $\sin\alpha = 1\times10^{-4}$ ($ 1\times10^{-5}$).
Here we have scanned $m_H$ for $2$ GeV $\leq m_H \leq 10$ GeV, 
 which is enough to find mass range of dark matter.
The lower bound on $m_H$ we took is due to the following reason.
The radiative correction to $\lambda_s$ from one loop diagram propagating $N_R$ is
 $\delta\lambda_s \simeq -(1/4\pi^2)\sum Y_R^4 $.
Provided that the largest coupling of $Y_R$ is the order of unity, 
as we often think from the viewpoint of ``naturalness'',
then we have $\delta\lambda_s = \mathcal{O}(-0.01)$. 
Thus, by considering such radiative corrections,
 $\lambda_s = 0.01$ seems to be a sensible lower value
 and its corresponding Higgs boson mass estimated from Eq.~(\ref{mass:H}) 
 is about $2$ GeV.
We show excluded regions by direct dark matter searches,
 the CREEST-II~\cite{Angloher:2014myn}, the CDMSlite~\cite{Agnese:2015nto} and
 the LUX~\cite{Akerib:2015rjg,Akerib:2016vxi}.
For both red and blue points, one can see two groups of points;
 the upper group predicting a larger cross section with a nucleon and
 the lower group predicting a smaller cross section.
The former corresponds to the case of Eq.~(\ref{thresh}) and $m_{DM}^+$ in Eq.(\ref{reso})
 and the later does to the case of Eq.~(\ref{off-reso}) and $m_{DM}^-$ in Eq.(\ref{reso}).
The projected sensitivity of the SuperCDMS SNOLAB~\cite{Agnese:2016cpb} can cover 
 the predicted regions of $\sin\alpha > \mathcal{O}(10^{-5})$.

As we mentioned above, among the dominant annihilation into $X$ bosons pair,
 only t-(u-) channel $N_R$ exchange contribution gives a small s-wave mode of $\mathcal{O}(10^{-5})$ pb.
Hence, the bounds from dark matter indirect searches
 such as the Fermi-LAT~\cite{Ahnen:2016qkx} do not constrain
 a parameter region of interest in this model.
It has been pointed out that, in light of AMS-02 data~\cite{Aguilar:2013qda},
 a low mass region ($m_{DM} < \mathcal{O}(0.1)$ GeV) of dark matter annihilating
 into electrons is stringently constrained~\cite{Ibarra:2013zia}.
Even with such a small annihilation cross section,
 it is not trivial to confirm that this constraint is satisfied for a light mass region.
Thus, we restrict investigated dark matter mass range at GeV region in this paper,
 we might study such light region elsewhere.

\section{Summary} \label{summary}

Motivated by the Atomki anomalies and nonvanishing neutrino masses, we have considered 
a gauged $U(1)_{B-L}$ extended radiative seesaw model. We showed that the anomalies as well as 
the dark matter abundance and nonvanishing neutrino masses can be explained simultaneously. 

Requiring the decay of the $B-L$ gauge boson to be the origin of the Atomki anomalies, we showed 
that the model parameters can be determined or constrained. 
The resulting mass of the lightest right-handed neutrino dark matter is below about $3.4$ GeV and that of the 
lighter $Z_2$ even scalar is also below about $20$ GeV 
assuming the coupling to be smaller than the unity. However, such light particles must almost 
decouple from the SM due to small couplings. Therefore, the Higgs sector remains the SM-like, which is 
consistent with the LHC results.

We have also found that the relic dark matter abundance can be reproduced by the annihilation into $XX$. 
It further constrains the scalar mixing angle and the dark matter mass.
We have shown the consistent model parameter region with $\Omega_{N_R}h^2 \simeq 0.1$
 where the elastic scattering cross section of the DM particle off nuclei can be 
 below the current bound from the CRESST-II, the CDMSlite and LUX experiments. However, 
 the cross section is predicted within the reach of the expected sensitivity of Ge HV detector. 
Therefore, our dark matter candidate is in practice detectable even for an extremely small Higgs mixing angle $\sin\alpha$.

\begin{acknowledgements}
T.~S would like to thank Y.~Maeda for fruitful discussion on nuclear experiments.
We are grateful for M.~Aoki's valuable comments letting us notice an error in the previous calculation. 
This work is supported, in part, by JSPS KAKENHI Grants No.~15K17654 (T.~S)
 and No. 26400243 (O.~S.) and by the SUHARA Memorial Foundation (O.~S.).
\end{acknowledgements}

\appendix

\section{Amplitude}

We give explicit formulas of the invariant amplitude 
 squared for the pair annihilation processes 
 of the RH neutrinos. 

\subsection{Annihilation into $X X$}

${\cal M}_1 $ denotes the amplitude by the $s$-channel Higgs bosons $h$ and $H$ exchange,
 while ${\cal M}_2 $ does that for the $t(u)$-channel $N$ exchange diagram.
\begin{align}
& \overline{|{\cal M}|}^2 = \overline{|{\cal M}_1+{\cal M}_2|}^2 , \\
& \overline{|{\cal M}_1|}^2
 = m_N^2 q_{B-L}^2 g_{B-L}^2 \left|\sin^2\alpha \frac{1}{s-m_h^2+i m_h \Gamma_h} 
  + \cos^2\alpha \frac{1}{s-m_H^2+i m_H \Gamma_H} \right|^2  
  \nonumber \\
  & \times(s-4m_N^2) \left( 1 +\frac{1}{2 m_X^4}\left(\frac{s}{2}- m_X^2\right)^2 \right), \\
& \int\frac{d \cos\theta}{2}\overline{|{\cal M}_2|}^2
 = 
\frac{32 g_{B-L}^4 q_{B-L}^4}{m_X^4 }\nonumber \\
& \left( \frac{ 4 m_N^4 s \left(s-4 m_X^2\right)+4 m_N^2 m_X^2 \left(4 m_X^2-s\right) \left(m_X^2+s\right)-m_X^4 \left(4 m_X^4+s^2\right) 
}{\sqrt{s-4 m_N^2} \sqrt{s-4 m_X^2} \left(s-2 m_X^2\right)}
\right. \nonumber \\
& \left. \times 
\ln\left[\frac{s-2 m_X^2-\sqrt{s-4 m_N^2} \sqrt{s-4 m_X^2}}{s-2 m_X^2+\sqrt{s-4 m_N^2}\sqrt{s-4 m_X^2}}\right]  \right. \nonumber \\
& \left. +\frac{2 m_N^4 \left(8 m_X^4-8 m_X^2 s+s^2\right)+m_N^2 m_X^4 \left(4 m_X^2+s\right)-2 m_X^8 }{\left(m_N^2 \left(s-4 m_X^2\right)+m_X^4\right)} \right) ,\\
& \int\frac{d \cos\theta}{2} \overline{( {\cal M}_1 {\cal M}_2^* + c.c )} \nonumber \\
 = &   \frac{8 \sqrt{2} g_{B-L}^4 q_{B-L}^4 v_s \lambda  m_N }{m_X^4 \sqrt{s-4 m_N^2} \sqrt{s-4 m_X^2} }
 \left( \frac{-\sin\alpha\left(m_h^2-s\right)}{\Gamma_h^2 m_h^2+\left(m_h^2-s\right)^2}+\frac{\cos\alpha \left(m_H^2-s\right)}{\Gamma_H^2 m_H^2+\left(m_H^2-s\right)^2} \right)
 \nonumber \\
& \times \left(\sqrt{s-4 m_N^2} \left(4 m_X^4-2 m_X^2 s+s^2\right) \sqrt{s-4 m_X^2}
 \right. \nonumber \\
& \left. +2 \left(m_N^2 \left(8 m_X^4-4 m_X^2 s+s^2\right)-2 m_X^6\right) 
\log \left[\frac{s-2 m_X^2-\sqrt{s-4 m_N^2} \sqrt{s-4 m_X^2}}{s-2 m_X^2+\sqrt{s-4 m_N^2} \sqrt{s-4 m_X^2}}\right]
\right) ,
\end{align}
 where $\theta$ is the scattering angle in the center of mass frame.

\subsection{Annihilation into $H H$}

${\cal M}_1 $ denotes the amplitude by the $s$-channel Higgs bosons $h$ and $H$ exchange,
 while ${\cal M}_2 $ does that for the $t(u)$-channel $N$ exchange diagram.
\begin{align}
& \overline{|{\cal M}|}^2 = \overline{|{\cal M}_1+{\cal M}_2|}^2 , \\
& \overline{|{\cal M}_1|}^2  =
   \frac{ \lambda_N^2}{4} (s-4 m_N^2 ) 
 \left|
  \frac{\sin\alpha}{s-m_h^2+i m_h \Gamma_h}\lambda_{hHH}
  - \frac{\cos\alpha}{s-m_H^2+i m_H \Gamma_H}\lambda_{HHH} \right|^2 , \\
& \int\frac{d \cos\theta}{2}\overline{|{\cal M}_2|}^2 \nonumber \\
  = &  \frac{\lambda_N^4}{2} \cos^4\alpha
  \left( -8 
  - \frac{4 (m_H^2 -4 m_N^2)^2}{m_H^4 - 4 m_H^2 m_N^2 + m_N^2 s}
  + \right. \nonumber \\
& \left.
  4\frac{ (6 m_H^4 - 32 m_N^4 + 16 m_N^2 s + s^2 - 
   4 m_H^2 (4 m_N^2 + s)) }{(s-2 mH^2)\sqrt{(s-4 m_N^2)(s-4 m_H^2)}}
   \ln\left[ \frac{s-2m_H^2+\sqrt{(s-4 m_N^2)(s-4 m_H^2)} }{s-2m_H^2-\sqrt{(s-4 m_N^2)(s-4 m_H^2)} } \right]
  \right) , \\
& \int\frac{d \cos\theta}{2} \overline{( {\cal M}_1 {\cal M}_2^* + c.c )} \nonumber \\
 = &  8\sqrt{2} \lambda_N^3 m_N \cos^2\alpha
  \left( -\frac{\sin\alpha (s-m_h^2) \lambda_{hHH}}{(s-m_h^2)^2+( M_h \Gamma_h)^2}
  + 
  \frac{\cos\alpha (s-m_H^2) \lambda_{HHH}}{(s-m_H^2)^2+ (M_H \Gamma_H)^2} \right) \nonumber \\
&  \left(
   1 + \frac{s-8m_N^2+2 m_H^2}{2\sqrt{(s-4m_N^2)(s-4 m_H^2)}} 
   \ln\left[\frac{s-2m_H^2+\sqrt{(s-4m_N^2)(s-4 m_H^2)}}{s-2m_H^2-\sqrt{(s-4m_N^2)(s-4 m_H^2)}}\right]
  \right) .
\end{align}
%

\bibliographystyle{apsrev}
\bibliography{biblio}

\end{document}